\documentclass[twocolumn]{aastex631}
\usepackage{appendix}
\usepackage{listings}
\usepackage{savesym} 
\savesymbol{tablenum}
\usepackage{siunitx}
\restoresymbol{SIX}{tablenum}

\begin{document}

\title{Identifying Telescope Usage in Astrophysics Publications: A Machine Learning Framework for Institutional Research Management at Observatories}

\author[0000-0003-2248-0941]{Vicente Amado Olivo}
\affiliation{Department of Computational Mathematics, Science, and Engineering, Michigan State University, East Lansing, MI 48824, USA}

\author[0000-0002-0479-7235]{Wolfgang Kerzendorf}
\affiliation{Department of Computational Mathematics, Science, and Engineering, Michigan State University, East Lansing, MI 48824, USA}
\affiliation{Department of Physics and Astronomy, Michigan State University, East Lansing, MI 48824, USA}

\author[0000-0002-4289-7923]{Brian Cherinka}
\affiliation{Space Telescope Science Institute, Baltimore, MD, 21218, USA}

\author[0000-0002-1560-5286]{Joshua V. Shields}
\affiliation{Department of Physics and Astronomy, Michigan State University, East Lansing, MI 48824, USA}

\author{Annie Didier}
\affiliation{Outrider, Golden, Colorado 80403, USA}

\author{Katharina von der Wense}
\affiliation{Department of Computer Science, University of Colorado Boulder, Boulder, CO 80309, USA}
\affiliation{Institute of Computer Science, Johannes Gutenberg University Mainz, 55128 Mainz, Germany}

\correspondingauthor{Vicente Amado Olivo}
\email{amadovic@msu.edu}

\begin{abstract}
Large scientific institutions, such as the Space Telescope Science Institute, track the usage of their facilities to understand the needs of the research community. Astrophysicists incorporate facility usage data into their scientific publications, embedding this information in plain-text. Traditional automatic search queries prove unreliable for accurate tracking due to the misidentification of facility names in plain-text. As automatic search queries fail, researchers are required to manually classify publications for facility usage, which consumes valuable research time. In this work, we introduce a machine learning classification framework for the automatic identification of facility usage of observation sections in astrophysics publications. Our framework identifies sentences containing telescope mission keywords (e.g., Kepler and TESS) in each publication. Subsequently, the identified sentences are transformed using Term Frequency-Inverse Document Frequency and classified with a Support Vector Machine. The classification framework leverages the context surrounding the identified telescope mission keywords to provide relevant information to the classifier. The framework successfully classifies usage of MAST hosted missions with a 92.9\% accuracy. Furthermore, our framework demonstrates robustness when compared to other approaches, considering common metrics and computational complexity. The framework's interpretability makes it adaptable for use across observatories and other scientific facilities worldwide.
\end{abstract}

\section{Introduction}\label{sec:intro} The global astronomical community identifies the scientific output of telescope facilities, which is essential to understand the needs of the community, effectively allocate resources, and plan current and future facility management. A primary metric for measuring this scientific output is tracking the usage of telescope facilities in scientific publications \citep[see, e.g.,][]{grothkopf_use_2018}. However, simple programmatic attempts at identifying telescope usage based on the appearance of mission keywords fail because the descriptions of telescope usage are embedded in an unstructured way throughout the plain-text of publications. Consequently, for now, researchers are required to manually identify publications and leverage specific subject knowledge needed for accurate identification. 

\begin{figure*}[t!]
\centering
\includegraphics[width=\textwidth]{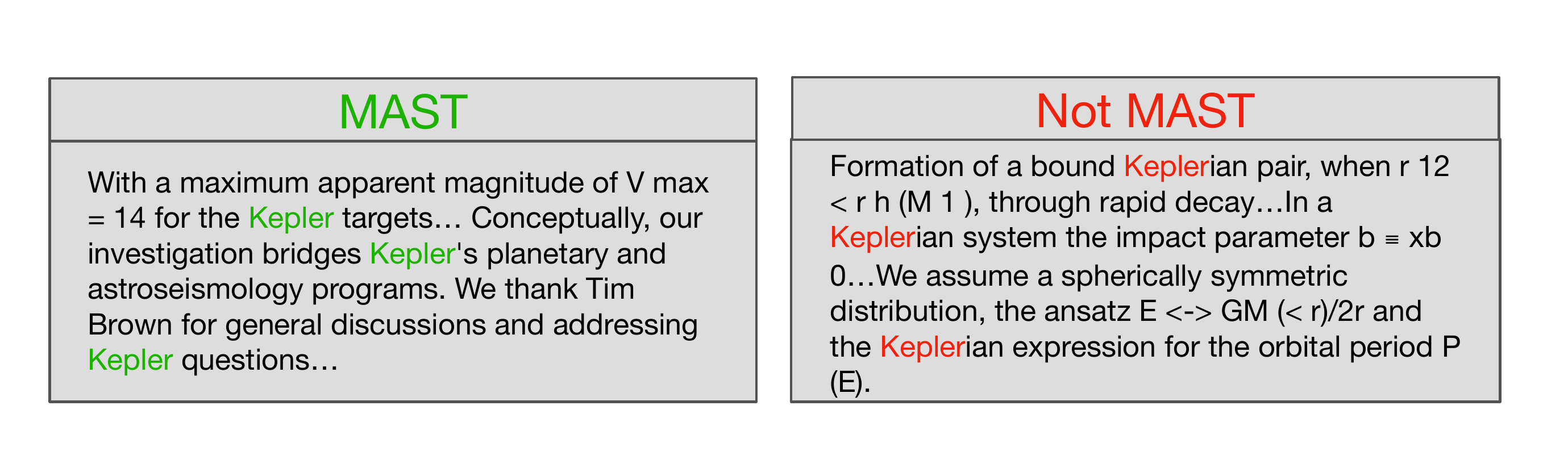}
\caption{The filtered dataset comprises sentences that include MAST hosted mission keywords. For example, when filtering publications for the space telescope 'Kepler' we identify three sentences shown in these two publications. Contextual cues guide the reader or classifier to distinguish that, on the left, one publication references the 'Kepler' space telescope and is a \textbf{MAST} publication, while on the right, the other publication discusses 'Keplerian physics' and the research does not utilize the Kepler space telescope.}
\label{fig:comparison_txt}
\end{figure*}

The exponential growth of the astronomical literature requires researchers to continually spend additional time manually classifying publications, a mundane task diverting researchers from their primary research \citep[see, for example,][]{kerzendorf_knowledge_2019, bornmann_growth_2021}. At the Space Telescope Science Institute (STScI), a small team of researchers queries the SAO/NASA Astrophysics Data System (NASA/ADS) \footnote{\url{https://ui.adsabs.harvard.edu/}} each month, dedicating part of their time to identifying and classifying as many relevant publications as possible, alongside their other responsibilities. For example, from 2021 to 2023, on average, $\sim{550}$ new publications per month were identified for classification. Given workforce capacity and resource limitations, on average, 14\% of these publications remain unclassified, contributing to a growing backlog (private communication, Brian Cherinka). The accurate and scalable tracking of telescope usage can potentially be met by automatic text classification.

Simple automatic solutions, such as full-text searches, cannot accurately distinguish the nuances that indicate telescope usage in astrophysics publications \citep{barnett_machine_2009}. Facility names (e.g., FUSE, Copernicus, and Kepler) can be misidentified owing to their lack of uniqueness and potential overlap with common words (see Figure~\ref{fig:comparison_txt}). While the lack of structure in the plain-text of publications requires initially searching for telescope names, accurate classification requires leveraging the context to identify telescope usage.  

Text classification is a principle task in the field of natural language processing (NLP), with various recent advancements with the specific aim to allow for the inclusion of domain knowledge and feedback processes \citep[e.g., fine-tuning;][]{beltagy_scibert_2019}. Accordingly, much research has been directed at presenting efficient, generalizable, and accurate methodologies with more applications appearing recently in the scientific domain \citep{khadhraoui_survey_2022, farshid_danesh_text_2023}. As simple automatic solutions fail, the advancements of computational techniques, like machine learning, are leveraged for the task of identifying telescope or data usage from publications \citep[][hereafter C22]{chen_classification_2022}. 

In the domain of astrophysics, C22 presents a classification framework trained to categorize publications relevant to NASA/IPAC Extragalactic Database (NED) data products. Utilizing the Stanford NLP classifier, C22 identifies NED-relevant publications. Nevertheless, various limitations exist, specifically the resource-intensive data preprocessing and inclusion of less informative features. These limitations introduce challenges when adapting the classification framework presented in C22 for our objective of identifying telescope usage (refer to Section~\ref{sec:results} for detailed insights).

To overcome the limitations, we present a robust, interpretable, and scalable text classification framework using machine learning to automatically identify telescope usage in publications. We train and validate the text classification framework using labeled data from The Barbara A. Mikulski Archive for Space Telescopes (MAST)\footnote{\url{https://archive.stsci.edu/}} as a case study. 

In Section~\ref{sec:methods}, we introduce the data used in the binary text classification task and outline the data preprocessing workflow. Additionally, we discuss the training process for the selected classifiers along with their hyperparameter tuning. In Section~\ref{sec:results}, we present the performance of the selected classifiers, utilizing various commonly used metrics, which we draw comparisons with C22. Finally, in Section~\ref{sec:conclusion}, we draw conclusions from our findings and emphasize the effectiveness of the developed classification framework in correctly tracking the usage of MAST hosted missions.

\begin{figure*}[t]
\centering
\includegraphics[width=0.9\textwidth]{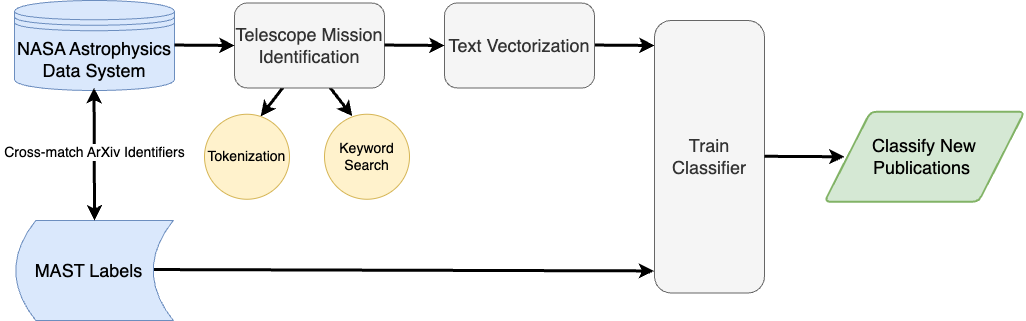}
\caption{Our classification framework, presented here, follows a supervised training structure. First we identify relevant sentences in each publication containing mission keywords (e.g., Kepler and TESS) and then we vectorize and classify the text.}
\label{fig:mast_workflow}
\end{figure*}

\section{Methods} \label{sec:methods}
Our text classification framework is designed and trained to determine whether a publication contains usage of any MAST hosted mission. In collaboration with MAST scientists, we utilize manually classified data to train a supervised binary classifier, employing the labels \textbf{MAST} and \textbf{\textbf{Not MAST}} along with associated publication metadata in the training process. Figure~\ref{fig:mast_workflow} presents the framework, which identifies the observation sections in astrophysics publications by filtering the full-text for sentences containing mission keywords (e.g., MAST and TUES). The identified observation sections are numerically represented using Term Frequency-Inverse Document Frequency (TF-IDF) for classifier training. We assess various classifiers' performance classifying the numerically transformed sentences into the two classes. Notably, the framework leverages contextual information around identified mission keywords to provide informative features for the classifier.

\subsection{Data}\label{subsec:data}
Historically, MAST researchers manually classified astrophysics publications from NASA/ADS. Specifically, they classified \num{32518} publications from 1996 to 2022, that were identified by querying NASA/ADS for publications containing one or more of the specific keywords listed in Appendix~\ref{sec:keywords}. Publications are labeled either as \textbf{MAST} or \textbf{Not MAST} based on the presence or absence of MAST hosted mission usage (see Appendix~\ref{sec:MAST} for details on the classification process at MAST). To be exact, there are \num{10638} \textbf{MAST} publications and \num{21881} \textbf{Not MAST} publications in the database. Additionally, relevant metadata (e.g., arXiv identifier\footnote{\url{https://info.arxiv.org/help/arxiv_identifier.html}}, bibcode, mission, and year) is aggregated along with the label.

To compile a dataset to train, validate, and test a supervised text classifier, we first obtained the full-text publications from the Semantic Scholar Open Research Corpus (S2ORC)\footnote{\url{https://allenai.org/data/s2orc}}. S2ORC is an open-access collection of scientific publications that allows for bulk access to full-text articles developed by the Allen Institute for AI \citep{lo_s2orc_2020}. We extend the metadata provided by MAST scientists by leveraging the full-text of the manually labeled astrophysics publications. The full-text contains informative features, including the MAST hosted mission usage description, which is not available in the metadata.

We then cross-matched the astrophysics publications available in S2ORC with the metadata from the labeled dataset from MAST, utilizing the available arXiv IDs to identify the full-text. The cross-matching resulted in a total of \num{14808} publications in our corpus (i.e., the collection of publications) with full-text data available, comprising \num{7664} \textbf{MAST} and \num[]{7144} \textbf{Not MAST} publications. The balanced subset was utilized to train, validate, and test a classification model. 

\subsection{Tokenization and Transformation}\label{subsec:tokens} We began our data preprocessing by splitting each publication in our corpus into smaller pieces, specifically by sentence, using the sentence tokenizer from the \textsc{Natural Language Toolkit} library in Python \citep{bird_nltk_2004}. Next, we parsed the text to identify the sentences containing specific mission keywords (e.g., Kepler, K2, MAST, TESS, etc.). 

\begin{table*}[t!]
\centering
\small
\begin{tabular}{|c|c|c|c|}
\hline
\# of Sentences &  Support Vector Machine & Random Forest & Multilayer Perceptron \\
\ before, after & & & \\
\hline
0,0 & 0.929 & 0.905 & 0.908 \\
2,0 & 0.919 & 0.899 & 0.911 \\
0,2 & 0.920 & 0.896 & 0.912 \\
2,2 & 0.915 & 0.897 & 0.910 \\
4,2 & 0.904 & 0.888 & 0.904 \\
2,4 & 0.904 & 0.894 & 0.902 \\
4,4 & 0.899 & 0.884 & 0.904 \\
Full-text & 0.815 & 0.814 & 0.848 \\
\hline
\end{tabular}
\caption{The table displays the performance of three classifiers (Support Vector Machine (SVM), Random Forest, and Multi-layer Perceptron) as the number of input sentences for training varies, assessing the required context for classification. The Support Vector Machine classifier, trained exclusively on sentences containing mission keywords, achieves a successful accuracy of 92.9\% on the test set}
\label{table:mast_table}
\end{table*}

We experimented varying the sentences surrounding a mission keyword identified in the full-text as input. The context varied from zero sentences either preceding or following a sentence containing an identified mission keyword to the entire full-text of the publication. Table~\ref{table:mast_table} presents the varied number of sentences surrounding a mission keyword utilized in training the classifiers. The context words surrounding a mission keyword included in each class differ and inform the classifier differently for classification of publications. The preliminary testing found that exclusively utilizing the sentences containing keywords generally leads to the highest performance. Notably, the restricted context window enables the model to better distinguish between the two classes. Therefore, we chose to limit classifier training to exclusively the sentences containing mission keywords in the classification framework. 

Figure~\ref{fig:comparison_txt} displays two publications filtered for a mission keyword and the context words included in the identified sentences. The classification is informed by the contextual words around a mission keyword.

We utilize TF-IDF to numerically represent the identified relevant sentences for training \citep{sparck_jones_statistical_1972, luhn_statistical_1957}. We utilize the TF-IDF formulation presented by the \textsc{Scikit-learn} TF-IDF vectorizer\footnote{\url{https://scikit-learn.org/stable/modules/generated/sklearn.feature_extraction.text.TfidfVectorizer.html}} \citep{pedregosa_scikit-learn_2011}:

\begin{equation}
\text{TF-IDF}(t, d) = \text{TF}(t, d) \times \text{IDF}(t)
\end{equation}

\begin{equation}
\text{TF}(t, d) = \frac{\text{count}(t, d)}{\text{maxCount}(d)}
\end{equation}

\begin{equation}
\text{IDF}(t) = \log \left( \frac{1+n}{\text{DF}(t) + 1} \right)+1
\end{equation}

Where $t$ represents a term (word) in the document $d$ across the total documents ($n$). The $\text{count}(t, d)$ indicates the number of times that term $t$ occurs in the document $d$, while $\text{maxCount}(d)$ denotes the total number of terms in the document $d$ and $\text{DF}(t)$ is the number of documents containing term $t$ across all documents in the set. The final TF-IDF score for each term $t$ in each document $d$ is given by $\text{TF-IDF}(t, d)$.

TF-IDF uses a vocabulary to construct the numerical transformation of each publication by considering all words present across publications. To optimize our classification framework, we tuned the TF-IDF vectorizer's hyperparameters by adjusting the vocabulary size, excluding terms appearing in less than \num[]{0.001} and more than \num[]{0.999} of the publications. We also used an ngram range of (1, 2) to include two-word sequences (e.g., white dwarf), which are relevant in the domain of astrophysics. This reduced the default vocabulary from over \num{40620} unique words to \num{29605} words. Consequently, the chosen minimum and maximum document frequency thresholds may lead to certain rare and ubiquitous contextual words being omitted from the final TF-IDF vector representations. Each publication is transformed into a vector using the constructed vocabulary, which resulted in each TF-IDF vector representation being $1 \times \num{29605}$. The hyperparameter tuning reduced the computational complexity by reducing the vocabulary size and in turn reducing the vector representation size. TF-IDF is an interpretable representation technique as each word in a given text is directly mapped to a corresponding element in the vector representation based on the vocabulary \citep{sitikhu_comparison_2019}. The constructed vocabulary guides the model in recognizing informative contextual words surrounding mission keywords for classification.

\begin{figure*}[t!]
\centering
\includegraphics[width=0.75\textwidth]{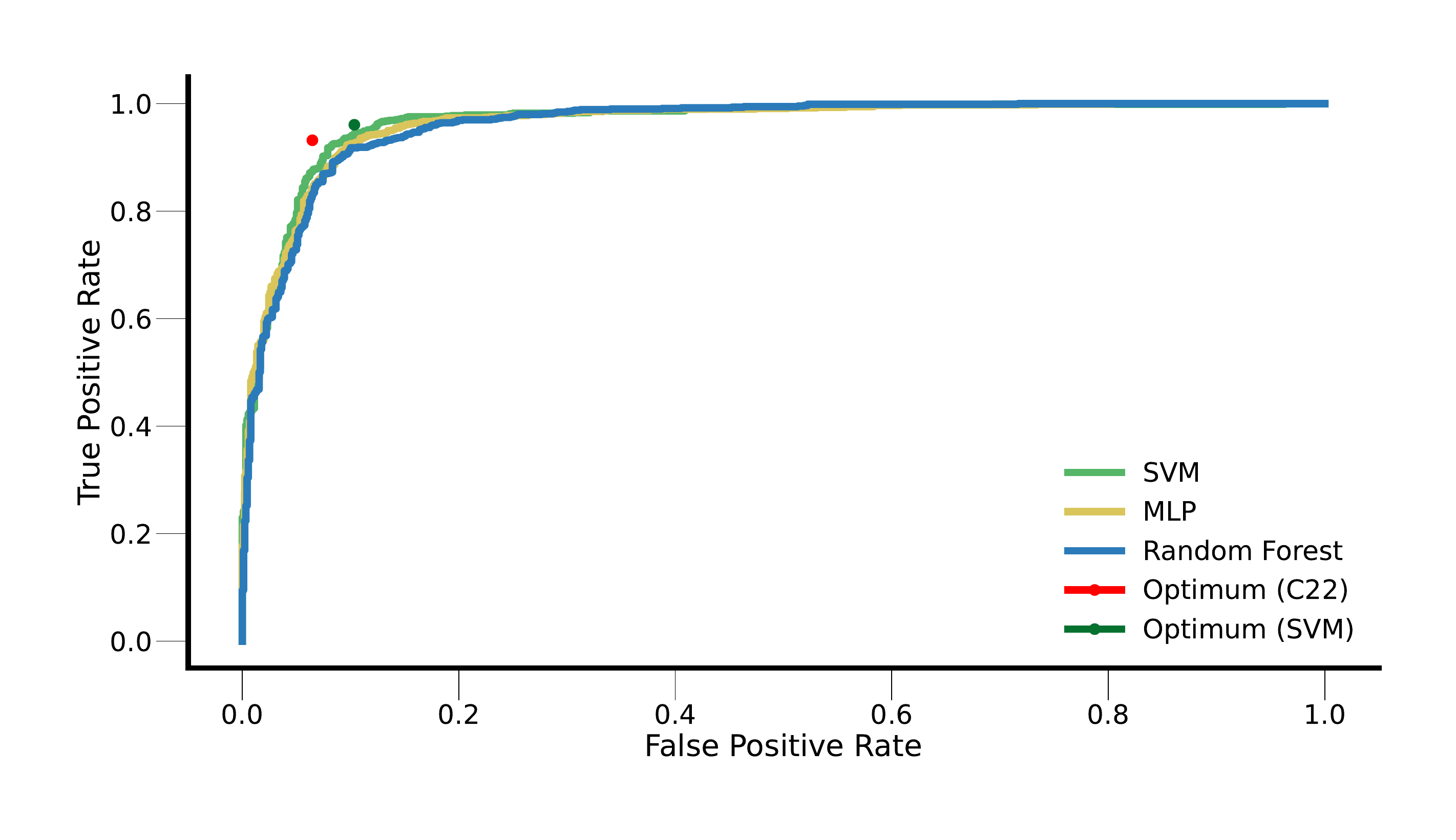}
\caption{We present the Receiver Operating Characteristic (ROC) curves displaying the performances of the SVM, Random Forest, and MLP classifiers. Notably, the SVM model trained exclusively on sentences containing mission keywords, has an Area Under the Curve (AUC) of 0.97. An AUC of \num[]{0.5} indicates that the model's ability to differentiate between the two classes is no better than random chance, while an AUC of one indicates the model's ability to perfectly distinguish between the two classes. The red point displays the optimum true-positive and false-positive rates from the classifier in paper C22, with a true-positive rate of \num[]{0.93} and a false-positive rate of \num[]{0.07}. In comparison, the dark green point displays that our SVM model has a higher true-positive rate of \num[]{0.96}. However, this improvement is accompanied by a slightly higher false-positive rate of \num[]{0.1}, indicating the inclusion of some irrelevant papers.}
\label{fig:mast_roc}
\end{figure*}

\subsection{Classifier Training}\label{subsec:classification} We chose to explore three classification algorithms: Support Vector Machine (SVM), Random Forest, and Multi-Layer Perceptron (MLP), implemented in Scikit-learn. For training and evaluation, we adopted a commonly used approach by randomly splitting our dataset of \num{14808} publications into training, validation, and test sets. The respective ratios were set at 60\%, 20\%, and 20\%, as determined to be close to optimal across various settings \citep[see, e.g.,][]{dobbin_optimally_2011}. 

We manually tuned the hyperparameters of the three classifiers. For the SVM, we experimented adjusting the class weight parameters to modify the weight assigned to the \textbf{MAST} publications class to decrease the false negative rate. In the case of the Random Forest, we configured the hyperparameters to include 500 estimators, set a verbosity level of 1, used 4 jobs, and specified a minimum of 10 samples per leaf. Similarly, we tuned the MLP utilizing an adaptive learning rate, 100 hidden layers, a ReLU activation function, and an \textsc{Adam} optimizer \citep[see, e.g.,][]{kingma_adam_2017}.

There is a trade-off when adjusting the class weight parameter to decrease the false negative rate—it will include more false positives, increasing the number of \textbf{Not MAST} publications classified as \textbf{MAST}. We prioritize minimizing the false negative rate by maximizing accurate classification of publications containing MAST hosted mission usage, while retaining an acceptable false positive rate currently set at 0.1 or lower. The threshold was determined based on its impact on the workload of MAST scientists at STScI. With the current monthly rate of approximately 550 new publications for classification and a false positive rate of 0.1, resulting in $\sim$\,55 incorrectly classified papers requiring manual review each month. The threshold of acceptable false positives is set by balancing both the classification accuracy and workload management, reducing the number of papers requiring manual review to a level that allows MAST scientists to efficiently conduct verification while maintaining time for their research.

\section{Results \& Discussion} \label{sec:results} The SVM classifier, trained exclusively on sentences containing mission keywords, achieves the highest accuracy of 92.9\%, up from 81.6\% when trained on the full text. Throughout the rest of the paper, when mentioning the SVM model we refer to the highest-performing SVM. In classification tasks the common metrics used are \citep[for more details, see][]{juba_precision-recall_2019}:  

$$\text{Recall} = \frac{TP}{(TP + FN)} $$
$$\text{Precision} = \frac{TP}{(TP + FP)}$$
$$\text{Accuracy} = \frac{TP+ TN}{(TP + FP + TN + FN)}$$. 

The SVM classifier achieves a recall of 96.0\%, a precision of 90.1\%, and an accuracy of 92.9\%. While the highest performing Random Forest classifier exhibits a recall of 92.0\%, a precision of 90.0\%, and an accuracy of 90.5\% and the highest performing MLP exhibits a recall of 92.0\%, a precision of 92.3\%, and an accuracy of 91.2\%.

\begin{figure*}[t!]
\centering
\includegraphics[width=\textwidth]{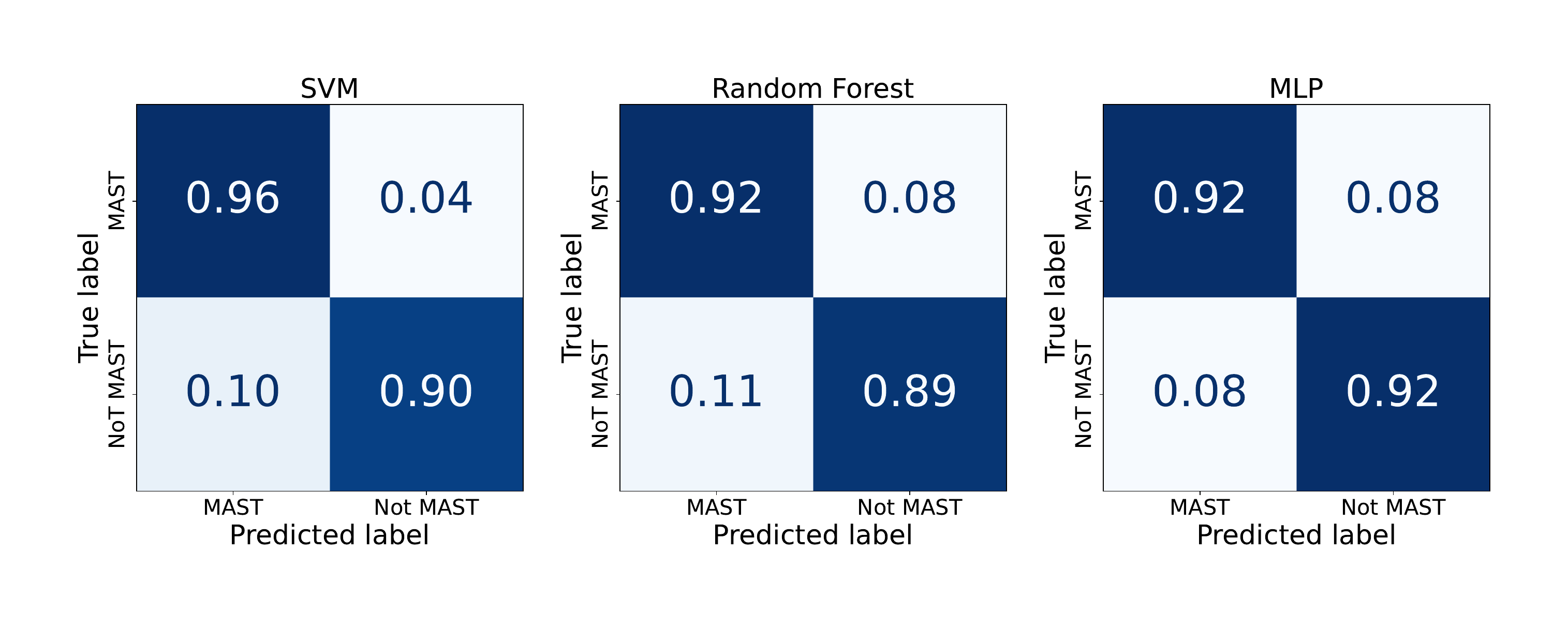}
\caption{A confusion matrix delineates the rates of predicted and true labels, enabling the evaluation of true and false predictions for each label in our binary classification \citep{maria_navin_performance_2016}. The SVM model exhibits a true-positive rate of 96\% in accurately predicting labeled \textbf{MAST} publications and a true-negative rate of 90\% in correctly identifying labeled \textbf{Not MAST} publications. In comparison, both the Random Forest model and MLP have lower true-positive rates, while the MLP has a slightly higher true-negative rate. }
\label{fig:mast_confusion}
\end{figure*}

To provide a more comprehensive view of the three classifiers' performances, we examine the Receiver Operating Characteristic (ROC) curve, depicted in Figure~\ref{fig:mast_roc}. An ROC curve with an area under the curve value of 0.97 signifies a high probability for the SVM model to distinguish between true positives and false positives. 

Figure~\ref{fig:mast_confusion} presents the confusion matrices for the three classifiers. The SVM model demonstrates a true-positive rate of 96\%, accurately predicting labeled \textbf{MAST} publications, and a true-negative rate of 90\%, correctly identifying labeled publications as \textbf{Not MAST} publications. Additionally, the Random Forest model achieves a true-positive rate of 92\% and a true-negative rate of 89\%, while the MLP exhibits a true-positive rate of 92\% and a true-negative rate of 92\%. Manual classification, while more accurate, is time-consuming. The objective of employing machine learning is to automate the process and ensure comprehensive identification of publications containing mission usage. These results above underscore how well the SVM model aligns with our goal of reducing the likelihood of excluding \textbf{MAST} publications by minimizing the false-negative rate.

We trained the classifier presented in C22 on our dataset, achieving a 90.2\% accuracy. We note the limitations of comparing our work to C22, as there are differences in data, preprocessing, and the specific problem addressed with the classifier. The C22 classifier is designed for the broader needs of NED, utilizing a comprehensive full-text feature extraction method. This approach is better suited for classifying observational data across a wide range of astronomical topics, where relying on a fixed set of keywords would be insufficient. In contrast, our classifier is specifically tailored to track usage of individual facilities, a more focused task that allows for a flexible and faster method.

Our approach, trained exclusively on sentences containing MAST-hosted mission keywords, offers several advantages. Unlike the resource-intensive feature extraction process in C22, which requires days of CPU time, our method is significantly faster and more scalable. Additionally, our classifier is less sensitive to changes in text structure or formatting, as it focuses on specific sentences in the text, compared to C22, which utilizes HTML files extracted from various astronomical journals as training data.
The NED classification framework's sensitivity to text structure nuances, such as document formatting or HTML templates, necessitates periodic retraining. Particularly, the alteration of a journal's HTML template would require retraining for new publications. As the volume of publications grows exponentially, our framework scales more effectively, requiring minimal retraining, which is primarily necessary when new mission keywords are added, making it more adaptable to evolving needs (see Appendix~\ref{sec:classifiers} for performance characteristics).
While the C22 classifier offers advantages for tasks across broad astronomical topics, our method provides a more targeted and efficient solution for tracking facility usage in MAST. This comparison highlights that the choice of classification method should be driven by the specific requirements of the problem at hand, with our approach being particularly well-suited for the needs of facility tracking.

\section{Conclusion}\label{sec:conclusion} We present a classification framework designed for large scientific institutions to effectively monitor the utilization of their scientific instruments and gain valuable insights into the evolving needs of the scientific community. The exponentially growing scientific literature has burdened MAST scientists who manually classify publications to track MAST hosted mission usage, leading to a persistent backlog of unclassified publications. 

Our classification framework leverages natural language processing techniques to automatically and accurately extract usage information. The presented framework workflow mirrors the interpretable classification methodology employed by MAST scientists, emphasizing the identification of sentences containing mission keywords within the plain-text of publications. Notably, the identification of relevant sentences containing mission keywords enhances accuracy by highlighting contextual information. Considering only sentences containing mission keywords improved classification accuracy from 81.6\% to 92.9\%, when compared to using the full-text publication (see Appendix~\ref{sec:fulltext} for further metrics regarding full-text).

The interpretability of TF-IDF enhances the framework's robustness by enabling direct evaluation of classifier performance at the word level. The explicit mapping of context words to elements of the vector representations facilitates the transparent assessment of the framework's effectiveness across various observatories. The integration of this classification model into the MAST workflow promises to alleviate manual classification and assist scientific institutions, such as STScI, in identifying the evolving needs of the global astronomical community. Future work will refine publication tracking for observatories, incorporating mission-specific and publication-type information to customize the framework for diverse observatory needs and enhance resource utilization tracking. 

\vspace{0.25cm}

\textbf{Acknowledgments}
We gratefully acknowledge funding support for Vicente Amado Olivo from the Space Telescope Science Institute. Additionally, we acknowledge the use of the SAO/NASA Astrophysics Data System API\footnote{\url{https://ui.adsabs.harvard.edu/help/api/}} service and extend our gratitude to the anonymous referee for the insightful comments that have improved this publication.


\textbf{Contributor Roles}
\begin{enumerate}
    \item Conceptualization: Wolfgang Kerzendorf, Brian Cherinka
    \item Data Curation: Brian Cherinka, Vicente Amado Olivo
    \item Formal Analysis: Vicente Amado Olivo
    \item Funding Acquisition: Wolfgang Kerzendorf 
    \item Investigation: Vicente Amado Olivo, Brian Cherinka, Wolfgang Kerzendorf, Katharina von der Wense, Annie Didier 
    \item Methodology: Vicente Amado Olivo, Wolfgang Kerzendorf, Katharina von der Wense, Annie Didier
    \item Project Administration: Wolfgang Kerzendorf
    \item Resources: Wolfgang Kerzendorf 
    \item Software: Vicente Amado Olivo
    \item Supervision: Wolfgang Kerzendorf
    \item Validation: Vicente Amado Olivo
    \item Visualization: Vicente Amado Olivo
    \item Writing - original draft: Vicente Amado Olivo, Brian Cherinka 
    \item Writing - reviewing \& editing: Brian Cherinka, Wolfgang Kerzendorf, Katharina von der Wense, Joshua V. Shields, Annie Didier
\end{enumerate}

\bibliography{MAST_Paper.bib}{}
\bibliographystyle{aasjournal}
\appendix

\vspace{-0.5cm}
\section{MAST Keywords}\label{sec:keywords}
Full list of keywords used by MAST Scientists to search for candidate MAST publications.

\begin{center}
\begin{tabular}{p{0.45\textwidth} p{0.45\textwidth}} 
    \begin{enumerate}
        \item PAN STARRS
        \item PANSTARRS
        \item PS1
        \item Pan-STARRS
        \item Pan-STARRS-1
        \item Pan-STARRS1
        \item PanSTARRS-1
        \item PanSTARRS1
        \item WUPPE
        \item UIT
        \item TESS
        \item TUES
        \item Berkeley Extreme and Far-UV Spectrometer
        \item Berkeley Spectrometer
        \item Transiting Exoplanet Survey Satellite
        \item Tubingen Ultraviolet Echelle Spectrometer
        \item Ultraviolet Imaging Telescope
        \item International Ultraviolet Explorer
        \item Interstellar Medium Absorption Profile Spectrograph
        \item Wisconsin Ultraviolet Photo-Polarimeter Experiment
        \item Orbiting Retrievable Far and Extreme Ultraviolet Spectrometers
    \end{enumerate} 
    &
    \begin{enumerate}
        \setcounter{enumi}{21} 
        \item AIDA
        \item BEFS
        \item Copernicus
        \item EUVE
        \item Extreme Ultraviolet Explorer
        \item FUSE
        \item Far Ultraviolet Spectroscopic Explorer
        \item GALEX
        \item Galaxy Evolution Explorer
        \item HUT
        \item Hopkins Ultraviolet Telescope
        \item IMAPS
        \item IUE
        \item 10.17909
        \item K2
        \item KTWOCANDELS
        \item Kepler
        \item MAST
        \item OAO-3
        \item ORFEUS
    \end{enumerate} 
\end{tabular}
\end{center}

\vspace{3cm}

\section{MAST Manual Classification Process}\label{sec:MAST}
\raggedright
In the following we describe, the manual classification process utilized by the MAST scientists. 

\subsection{Keyword Search}
\raggedright
The MAST archive scientists classify only a subset of papers from NASA/ADS. First, a full-text keyword search is done for all papers that contain any mission or MAST-related keywords, e.g. MAST, Kepler, TESS (see the full list in Appendix~\ref{sec:keywords}). Flagship missions like the Hubble Space Telescope Mission and James Webb Space Telescope Mission are separately classified. This search is done periodically, aggregated by month, currently producing on average roughly \num{550} publication results each month for further human verification.

\subsection{Classification Criteria}
\raggedright
Next, the publications are classified and labeled with the following information:
\begin{itemize}
    \item \textbf{Mission}: the relevant mission name (i.e., TESS, Kepler) 
    \item \textbf{Paper type}: which paper category the publication fits into (e.g., a "science paper" or a "mention")
    \item \textbf{Ignored}: whether the publication should be ignored altogether
\end{itemize}

\raggedright
Publications marked as \textbf{Ignored} are typically papers where the matched keyword refers to something unrelated to astronomical mission data at MAST. For example, the \textbf{MAST} acronym keyword can reference "Mega Amp Spherical Tokamak," rather than "Mikulski Archive for Space Telescopes," or the \textbf{K2} mission keyword can reference a variable in a mathematical equation, \(k^2\), rather than NASA's K2 mission. The contaminations in the initial keyword search are due to limitations with the NASA/ADS keyword search and must be accounted for during classification. See the \textbf{Not MAST} box in \textbf{Figure~\ref{fig:comparison_txt}} for an additional example of a keyword mismatch.

For binary classification, the archive scientists created a new column indicating for each paper whether it is a MAST paper or not. Any paper with at least a mention of a relevant keyword is considered a MAST paper.  All papers that were classified as "ignores" are considered as \textbf{Not MAST} papers.

\section{Performance of Classifiers}\label{sec:classifiers}
\subsection{Code Availability}
\raggedright
The code and data is open-source and is available here (\url{https://zenodo.org/records/14014142}) as a Zenodo repository. 

\subsection{Profiling}
\raggedright
The performance of our classification framework was profiled for execution time utilizing Python's timing functionality, with all measurements repeated over 7 runs to calculate mean times and standard deviations. We profiled the best-performing SVM, Random Forest, and MLP classifiers. The data processing steps, comprising tokenization, keyword search, and TF-IDF computation, exhibited the following execution times per run: tokenization required \num{4} minutes \num{8} seconds ($\pm$ \num{694} ms), keyword search took \num{12.3} seconds ($\pm$ \num{56.7} ms), and TF-IDF computation lasted \num{5.5} seconds ($\pm$ \num{27.7} ms).

The Random Forest classifier trains in \num[]{11.7} seconds ($\pm$ 47.4 ms) per run with an accuracy of 90.5\% on the test set. In comparison, the MLP trains in \num{210} seconds ($\pm$ 29.6 s) with an accuracy of 90.8\%. The SVM as discussed in Section~\ref{sec:results} is the classifier used in the framework as the accuracy of 92.9\% is the highest of the tested classifiers, while the training time is \num{40.3} seconds ($\pm$ 218 ms) per run.


\section{SVM Full-text Metrics}\label{sec:fulltext}
ROC curve and confusion matrix for an SVM model trained on the full-text publications.
\begin{figure*}[!ht]
\centering
\includegraphics[width=0.75\textwidth]{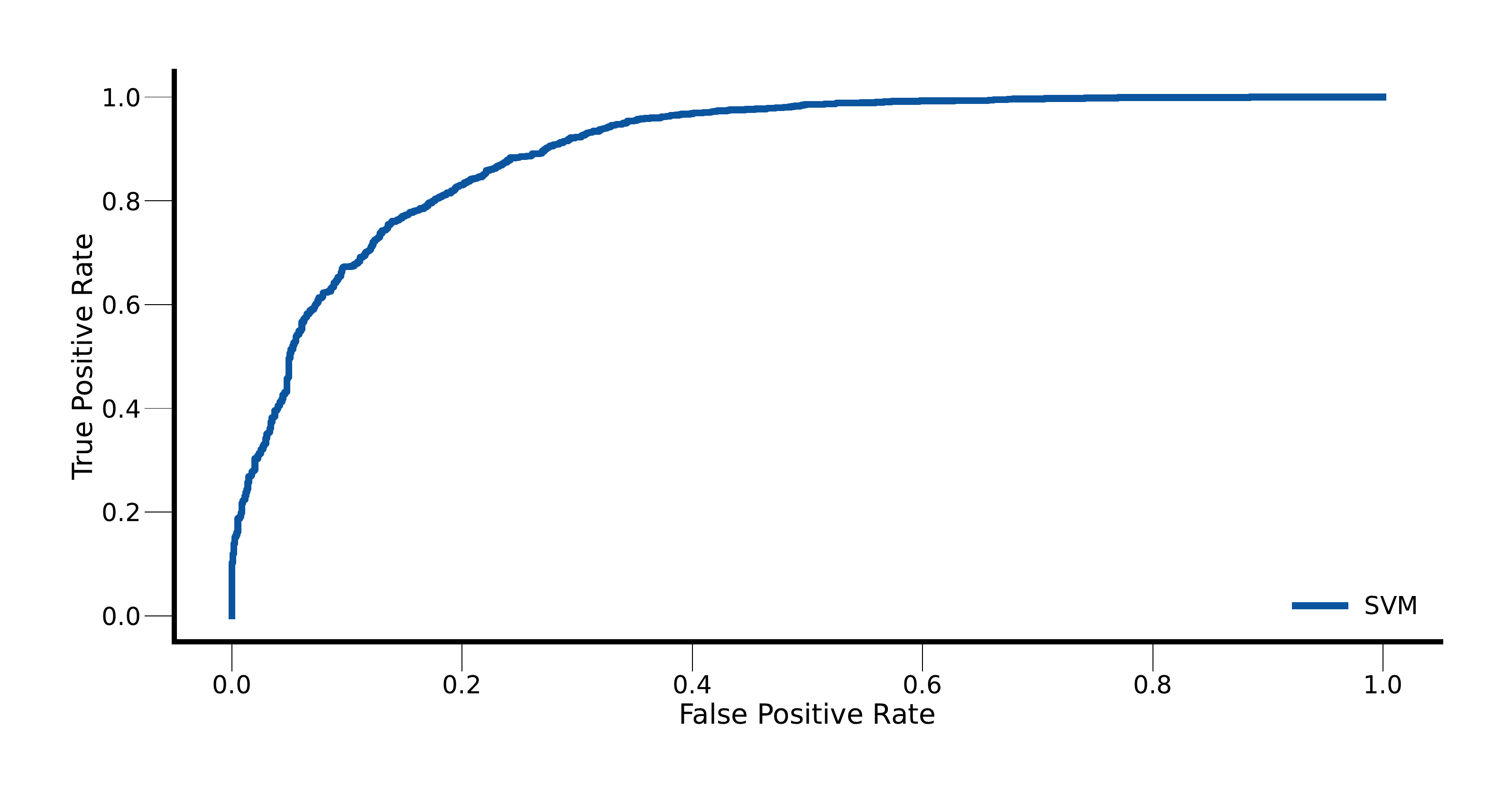}
\caption{The ROC curve of the SVM model trained on the full-text publications has an Area Under the Curve (AUC) of 0.90 compared to 0.97 when identifying the sections of the publication containing keywords.}
\end{figure*}

\begin{figure*}[!ht]
\centering
\includegraphics[width=0.5\textwidth]{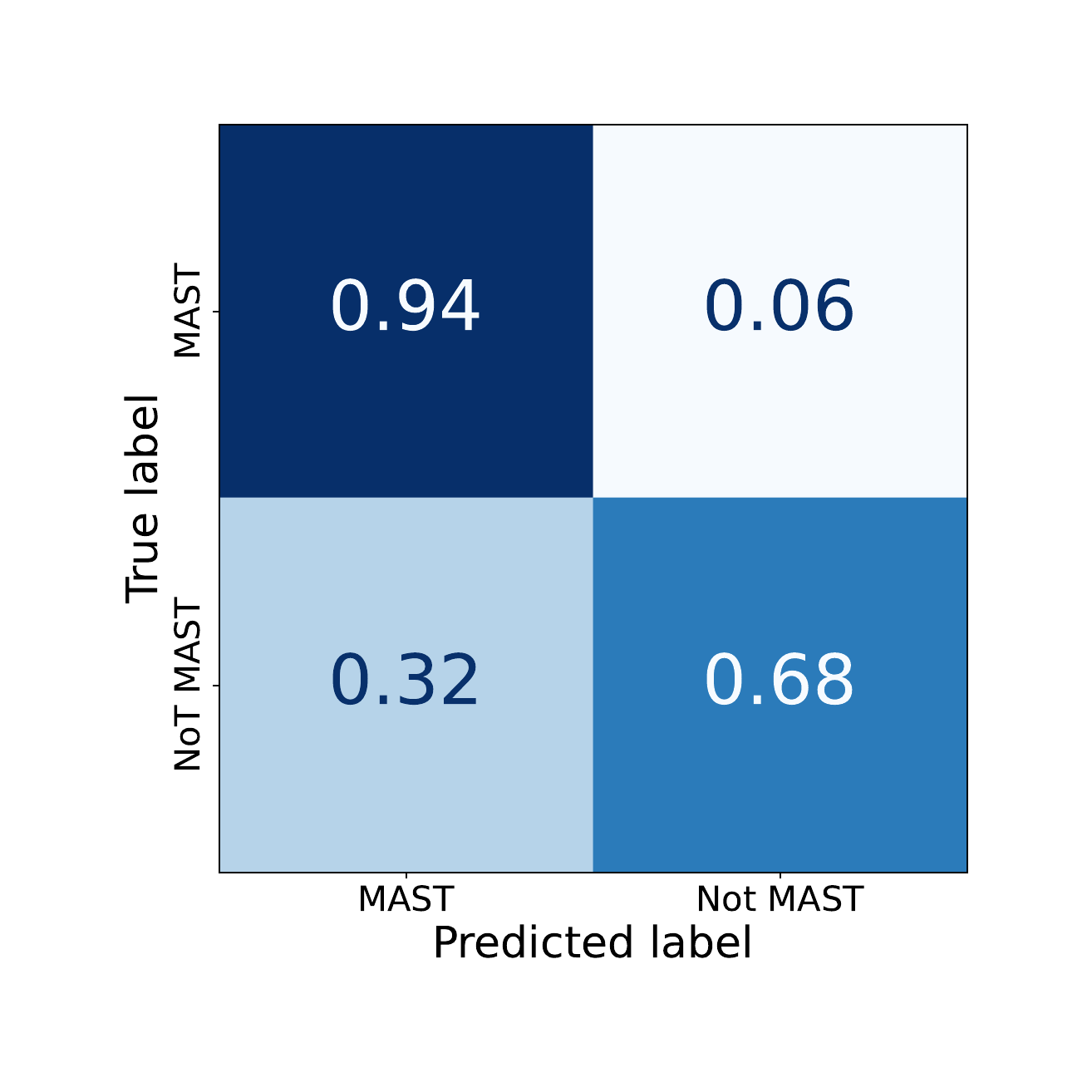}
\caption{The SVM model using the full-text publications, instead of identifying the sections with key-words, exhibits a true-positive rate of 94\% in accurately predicting labeled \textbf{MAST} publications and a true-negative rate of 68\% in correctly identifying labeled \textbf{Not MAST} publications.}
\end{figure*}

\end{document}